\documentclass{kapproc} 

\usepackage{procps} 

\usepackage[dvips]{graphicx}

\upperandlowercase

\normallatexbib 

\newcommand\beq{\begin{equation}}
\newcommand\eeq{\end{equation}}
\newcommand\beqa{\begin{eqnarray}}
\newcommand\eeqa{\end{eqnarray}}

\newcommand{\ds}[1]{#1 \hspace{-0.5em}/}  
\newcommand\bzeta{\mbox{\boldmath$\zeta$}}

\newcommand\btau{\mbox{\boldmath$\tau$}}
\newcommand\btheta{\mbox{\boldmath$\theta$}}
\newcommand\bk{{\bf k}}
\newcommand\bq{{\bf q}}

\def\sla{\slash{\!\!\!} }
\newcommand{\vp}{\mbox{\boldmath $p$}}

\begin{document}



\articletitle[Possibility of color magnetic
 superconductivity]{Possibility of color magnetic\\ superconductivity}












\author{Toshitaka Tatsumi}
\affil{Department of Physics, Kyoto University, Kyoto 606-8502, Japan}
\email{tatsumi@ruby.scphys.kyoto-u.ac.jp}

\author{Tomoyuki Maruyama}
\affil{College of Bioresouce Sciences, Nihon University, 
         Fujisawa, 252-8510, Japan}
\email{tomo@brs.nihon-u.ac.jp}

\author{Eiji Nakano}
\affil{Department of Physics, Tokyo Metropolitan University, 
         1-1 Minami-Ohsawa, Hachioji, Tokyo 192-0397, Japan}
\email{enakano@comp.metro-u.ac.jp}








\begin{abstract}
Two aspects of quark matter at high density are addressed: one is color
 superconductivity and the other is ferromagnetism. We are
 mainly concerned with the latter and its relation to color
 superconductivity, which we call {\it color magnetic superconductivity}.
 The relation of ferromagnetism and chiral symmetry restoration is also
 discussed. 
\end{abstract}


\section{Introduction}

Nowadays it is widely accepted that there should be realized various phases of
QCD in temperature ($T$) - density ($\rho_B$) plane. When we emphasize
the low $T$ and high $\rho_B$ region, the subjects are sometimes called
physics of high-density QCD. The main purposes in this field should be 
to figure out the properties of phase transitions and new phases, and to
extract their symmetry breaking pattern and low-energy excitation modes
there on the basis of QCD. On the other hand, these studies have
phenomenological implications on relativistic heavy-ion collisions and
compact stars like neutron stars or quark stars.

In this talk we'd like to address magnetic properties of quark matter at
low temperature. We first discuss the ferromagnetic phase transition and
then a possibility of the coexistence of ferromagnetism (FM) and color
superconductivity (CSC). We also present an idea about how FM is related
to chiral symmetry.

CSC should be very popular and many people believe that it is robust due
to the Cooper instability even for small attractive quark-quark
interaction in color $\bar 3$ channel \cite{CSC1}. On the contrary, we are afraid
that FM has not been so familiar yet. So, we'd like to begin with a brief
introduction about our motivation for the study of FM.

Phenomenologically the concept of magnetism should be directly related
to the origin of strong magnetic field observed in compact stars \cite{MAG3}; e.g., it
amounts to $O(10^{12}$G) at the surface of radio pulsars. Recently a new
class of pulsars called magnetars has been discovered with super strong
magnetic field, $B_s\sim 10^{14 - 15}$G, estimated from the $P-\dot{P}$
curve \cite{MAG1, MAG2}. First observations  are indirect evidences for 
super strong magnetic field, but 
discoveries of some absorption lines stemming from the cyclotron
frequency of protons have been currently reported \cite{pop}. 

The origin of the
strong magnetic field has been a long standing problem since the first
pulsar was discovered \cite{MAG3}. A naive working hypothesis is the
conservation of the magnetic flux and its squeezing during the evolution
from a main-sequence progenitor star to a compact star, $B_s\propto R^{-2}$ with
$R$ being the radius \cite{gin}.   

\begin{table}[h]
\caption{Surface magnetic field and the radius of stars by the
 conservation of the magnetic flux. }
\centering
\begin{tabular*}{5cm}{@{\extracolsep{\fill}}lll}
\sphline
 ~& $B_S[{\rm G}]$ & $R [{\rm cm}]$\\ \sphline
{\rm Sun~(obs.)} & $10^3$ & $10^{10}$\\
{\rm Neutron star} & $10^{11}$ & $10^6$\\ \sphline
{\rm Magnetar} & $10^{15}$ & $10^4$
\end{tabular*}
\end{table}

The relation of the radius and the expected strength of the 
magnetic field is listed by the use of the hypothesis in Table.~1. 
Then, it looks to work well for explaining the strength of the magnetic field
observed for radio pulsars. However, it does not work for magnetars;
considering  the Schwatzschild radius, 
\beq
R_{\rm Sch}=2GM/c^2=4\times 10^5 [{\rm cm}]\gg 10^4 [{\rm cm}],
\eeq
for the canonical mass of $M=1.4M_\odot$, we are immediately led to a contradiction.

Since there should be developed hadronic matter inside compact
stars, it would be reasonable to consider a microscopic origin of such
strong magnetic field: ferromagnetism or spin polarization is one of the
candidates to explain it. To graspe a rough idea about how hadronic
matter can give such a super strong magnetic field rather easily, it should be
interesting to compare typical energy scales in some systems (see Table~2): 
the magnetic interaction energy is estimated as $E_{\rm mag}=\mu_iB$ with
the magnetic moment, $\mu_i=e_i/(2m_i)$.
\begin{table}[h]
\caption{Magnetic interaction energies $E_{\rm mag}$ for $10^{15}$G 
and the typical energy
 scales $E_{\rm typ}$ in electron, nucleon and quark systems.}
\centering
\begin{tabular*}{10cm}{@{\extracolsep{\fill}}lccc}
\sphline
 &electron & proton & quark\\ \sphline
$m_i[{\rm MeV}]$& 0.5 & $10^3$ & 1- 100\\
$E_{\rm mag}[{\rm MeV}]$~~~ & 5 - 6  &~~~$2.5\times 10^{-3}$~~~& $2.5\times 10^{-2} -
2.5$\\ \sphline
$E_{\rm typ}$ & {\rm KeV} & {\rm MeV} & {\rm MeV}
\end{tabular*}
\end{table}
Thus we can see $E_{\rm typ}\ll E_{\rm mag}$ for electrons, while
$E_{\rm typ}>E_{\rm mag}$ for nucleons or quarks. This simple
consideration may imply that strong interaction gives a feasible origin 
for the strong magnetic field. 
The possibility of
ferromagnetism in nuclear matter has been elaborately studied when the
pulsars were observed, but negative results have been reported so
far \cite{pand}. Here
we consider its possibility in quark matter from a different point of
view \cite{Tatsu}.

\section{What is ferromagnetism in quark matter?}

Quark matter bears some resemblance to electron gas interacting with the
Coulomb potential; the
gluon exchange interaction in QCD is similar to the electromagnetic
interaction in QED 
and color neutrality of quark matter corresponds to charge neutrality of
electron gas under the background of positively charged ions. It was
Bloch who first suggested a mechanism leading to ferromagnetism of
itinerant electrons \cite{blo, yoshi}. The mechanism is very simple but largely reflects
the Fermion nature of electrons. Since there works no direct interaction
between electrons as a whole, the Fock exchange interaction gives a
leading contribution; it can be represented as 
\beq
V_{\rm Fock}=-e^2\frac{1+\bzeta\cdot\bzeta'}{|\bk-\bq|^2},
\label{fock}
\eeq  
between two electrons with momenta, $\bk$ and $\bq$, and spin
polarizations, $\bzeta$ and $\bzeta'$, where the vector $\bzeta$
specifies the definite spin polarized state, e.g. $\bzeta=(0,0,\pm 1)$ for
spin up and down state. Then it is immediately conceivable that a most
attractive channel is the parallel spin pair, whereas electrons with
opposite polarizations gives null contribution.
This is nothing but a
consequence of the Pauli exclusion principle: electrons with the same
spin polarization cannot closely approach to each other, which
effectively avoid the Coulomb repulsion. On the other hand a polarized
state should give a larger kinetic energy by rearranging the two Fermi
spheres. Thus there is a trade-off between kinetic and interaction
energies, which leads to a {\it spontaneous spin polarization (SSP)} 
 or FM at some density
\footnote{FM does not necessarily accompany SSP in some cases with
internal degrees of freedom, as is seen in section 4.}
. One of the essential points we learned here is that 
we need no spin-dependent interaction at the original Lagrangian to see
SSP. We can see a similar phenomenon in dealing with nuclear matter
within the relativistic mean-field theory, where the Fock interaction
can be extracted by way of the Fierz transformation from the original
Lagrangian \cite{MaruTatsu}. 

Then it might be natural to ask how about in QCD. We list here some
features of QCD related to this subject. (1) the quark-gluon interaction
in QCD is rather simple, compared with the nuclear force; it is a gauge
interaction like in QED. (2) quark matter should be a color neutral
system and only the $\it exchange$ interaction is relevant like in
the electron system. (3) there is an additional flavor degree of freedom in
quark matter; gluon exchange never change flavor but it comes in through
the generalized Pauli principle. (4) quarks should be treated
relativistically, different from the electron system.

The last feature requires a new definition and formulation of SSP or FM in
relativistic systems since``spin'' is no more a good quantum number in
relativistic theories;
spin couples with momentum and its direction changes during the motion.  
It is well known that the Pauli-Lubanski vector $W^\mu$ is the four vector to represent the
spin degree of freedom in a covariant form,
\beq
W^\mu=-\frac{1}{4}\epsilon_{\mu\nu\rho\sigma}k^\nu\sigma^{\rho\sigma}.
\label{aaa}
\eeq
In the rest frame,
\beq
W^0=0, \qquad \frac{{\bf W}}{m}=\frac{1}{2}\gamma^5\gamma^0
\mbox{\boldmath$\gamma$}=\frac{1}{2}\mbox{\boldmath$\Sigma$},
\label{aac}
\eeq
where $
\mbox{\boldmath$\Sigma$}=\left(\begin{array}{cc}
\mbox{\boldmath$\sigma$} & 0 \\
0 & \mbox{\boldmath$\sigma$}
\end{array}\right)
$ in the usual basis.
For any space-like four vector $a$ orthogonal to $k$, $a^\mu k_\mu=0$,
we then have a property,
\beq
W\cdot a=-\frac{1}{2}\gamma_5\ds{a}\ds{k}.
\label{aab}
\eeq
By taking a 4-pseudovector $a^\mu$ s.t.
\beq
{\bf a}=\bzeta+\frac{\bk(\bzeta\cdot\bk)}{m(E_k+m)}, 
~a^0=\frac{\bk\cdot\bzeta}{m}~~~
\label{ac}
\eeq
with the axial vector $\bzeta$, we can see the operator 
\beq
P(a)=\frac{1}{2}(1+\gamma_5\ds{a})
\eeq
is the projection operator for the definite spin-polarized states;
actually $a^\mu$ is reduced to a three vector $(0, \bzeta)$
in the rest frame and we can allocate $\bzeta=(0,0,\pm 1)$ to spin ``up'' and ``down'' states.
Thus we can still use $\bzeta$ to specify the two intrinsic polarized states
even in the general Lorentz frame.

We briefly present a heuristic argument how quark matter becomes ferromagnetic by
the use of above definition \cite{Tatsu}.
The Fock exchange interaction, $f_{{\bf k}\zeta,{\bf q}\zeta'}$, between
two quarks is defined by 
\beq
 f_{{\bf k}\zeta,{\bf q}\zeta'}
=\frac{m}{E_k}\frac{m}{E_q}{\cal M}_{{\bf k}\zeta,{\bf q}\zeta'}.
\label{ce}
\eeq
${\cal M}_{{\bf k}\zeta,{\bf q}\zeta'}$ is the usual Lorentz
invariant matrix element and can be written in the lowest order as
\begin{eqnarray}
{\cal M}_{{\bf k}\zeta,{\bf q}\zeta'}
=g^2\frac{2}{9m^2}[2m^2-k\cdot q-m^2 a\cdot b]\frac{1}{(k-q)^2},
\label{cf}
\end{eqnarray}
where the spin dependent term renders 
\beqa
a\cdot b&=&-\frac{1}{m_q^2}\left[
-(\bk\cdot\bzeta)(\bq\cdot\bzeta')
+m^2\bzeta\cdot\bzeta'\right.\nonumber\\
&+&\left\{m(E_k+m)(\bzeta\cdot\bq)(\bzeta'\cdot\bq)
+m(E_q+m)(\bzeta'\cdot\bk)(\bzeta\cdot\bk)\right.\nonumber\\
&+&\left.\left.(\bk\cdot\bq)(\bzeta\cdot\bk)
(\bzeta'\cdot\bq)\right\}/(E_k+m)(E_q+m)\right].
\label{cg}
\eeqa
It exhibits a complicated spin-dependent structure arising from the
Dirac four spinor, while it is reduced to a simple form,
\beqa
-\frac{2}{9}g^2\frac{1+\bzeta\cdot \bzeta'}{({\bf
k}-{\bf q})^2}
\label{nr}
\eeqa
in the non-relativistic limit as in the electron system. Eq.~(\ref{nr})
clearly shows why parallel spin pairs are favored, while we cannot see
it clearly in the relativistic expression (\ref{cg}). We have
explicitly demonstrated that the ferromagnetic phase should be realized at 
relatively low density region \cite{Tatsu}.

\subsection{Relativistic ferromagnetism}

If we understand FM or magnetic properties of quark matter more deeply,
we must proceeds to a self-consistent approach, like Hartree-Fock
theory,  
beyond the previous perturbative argument. In ref.~\cite{MaruTatsu} we
have described how the axial-vector mean field (AV) and the tensor one 
appear as a consequence of the Fierz transformation within the
relativistic mean-field theory for nuclear matter, which is one of the
nonperturbative frameworks in many-body theories and corresponds to
the Hatree-Fock approximation. We also demonstrated
they are responsible to ferromagnetism of nuclear matter. An important
point obtained there is the ``condensation'' of AV.
\footnote{There appears no tensor mean field in QCD as a result of
chiral symmetry. So we, hereafter, only consider AV.} 

When we consider the non-vanishing AV in quark matter,
\beq
{\bf V} = -\gamma_5 \gamma_3 {\bf U}_A , ~{\bf U}_A//\hat z ,
\eeq
we see an interaction between quarks and AV,
\beq
H_{int}\propto \mbox{\boldmath$\sigma$}\cdot {\bf U}_A=\sigma_3U_A,~~~
U_A\geq 0,
\eeq
in a similar form to the magnetic interaction in QED. Then the quark
propagator in AV renders 
\beq
G^{-1}_A(p) = \sla{p}-m-\sla{\mu}+\gamma_5 \sla{U_A}.  
\label{propa}
\eeq
The poles of $G_A(p)$, $\det$$G^{-1}_A$($p_0$$=$$\epsilon_n$)$=$$0$,
give the single-particle energy spectrum:
\beqa
&& \epsilon_n=\pm \epsilon_\pm \\
  && \epsilon_{\pm}= \sqrt{{\bf p}^2+{\bf U}_A^2+m^2 \pm 2 
                        \sqrt{m^2 {\bf U}_A^2+({\bf p}\cdot {\bf U}_A)^2 }},
\label{spect}
\eeqa
where the subscript $\pm$ in the energy spectrum represents spin degrees of
freedom, and the dissolution of the degeneracy
corresponds to the {\it exchange splitting} of 
different ``spin'' states \cite{yoshi}. Actually it is reduced to a familiar form,
$
\epsilon_{\pm}=m+\frac{p^2}{2m}\pm  U_A,
$
in the non-relativistic limit, while 
\beq
\epsilon_{\pm}=\sqrt{p_t^2+(|p_z|\pm U_A)^2}
\label{relaspe}
\eeq
in the extremely relativistic limit, $m\rightarrow 0$. Note that $U_A$
only shifts the value of momentum in Eq.~(\ref{relaspe}), so that it
should be {\it redundant} in the massless case.

There are two Fermi seas for a given quark number with different volumes
due to the exchange splitting in the energy spectrum. 
The appearance of the rotation symmetry breaking term, $\propto {\bf
p}\cdot {\bf U}_A$ in the energy
spectrum (\ref{spect}) implies deformation of the Fermi sea: so
rotation symmetry is violated in the momentum space as well as the
coordinate space, $O(3)\rightarrow O(2)$. Accordingly the Fermi sea
of majority quarks exhibits a prolate shape ($ F^-$), while that 
of minority quarks an oblate shape ($F^+$) as seen Fig.~1
\footnote{On the contrary, the Fermi sea remains spherical in the
non-relativistic case \cite{yoshi}. It would be also interesting to
compare our results with those given in the different context \cite{Muther}.}
.
\begin{figure}[h]
\begin{center}
\includegraphics[width=10cm,clip]{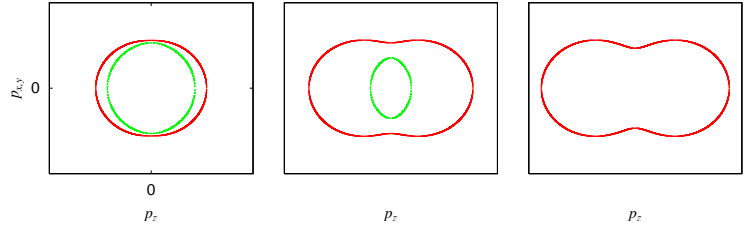}
\end{center}
\caption{Modification of the Fermi sea as $U_A$ is increased from left
 to right. The larger Fermi sea ($F^-$) takes a prolate shape, while
 the smaller one ($F^+$) an oblate shape for a given $U_A$. In the
 large $U_A$ limit (completely polarized case), $F^+$ disappears as 
in the right panel.}
\end{figure}

The mean spin polarization is then given by 
\beqa
{\bar s_z}=\frac{1}{2}\langle
\Sigma_z\rangle&=&-i\int_C\frac{d^4p}{(2\pi)^4}{\rm tr}
[\gamma_5\gamma_3G_A(p)]\\
&=&{\frac{1}{2}\left[\int_{F^+}\frac{d^3p}{(2\pi)^3} 
\frac{U_A+\beta}{\epsilon_+}+\int_{F^-}\frac{d^3p}{(2\pi)^3} 
\frac{U_A-\beta}{\epsilon_-}\right]}
\label{spol}
\eeqa
with $\beta=\sqrt{p_z^2+m^2}$, from which we can immediately see the
non-vanishing value of $U_A$ gives rise to spin polarization. 
Since the spin polarization is not necessarily measurable quantity, we'd
better to see another observable, the {\it magnetization}, which is
defined as the magnetic moment per unit volume and the
magnetic field directly couples with it.  
In QED the
magnetic field couples with quarks by way of the term, $\mu_q{\bar
q}\sigma_{\mu\nu}q F^{\mu\nu}$, with the Dirac magnetic moment
$\mu_q=e/(2m)$, and we can easily see that the
magnetization ${\bf M}$ is directed to the $z$ direction;
\beqa
M_3&=&-i\int_C\frac{d^4p}{(2\pi)^4}{\rm tr}
[\gamma_0\gamma_5\gamma_3G_A(p)]\\
&=&{\frac{1}{2}\left[-\int_{F^+}\frac{d^3p}{(2\pi)^3} 
\frac{m}{\beta}+\int_{F^-}\frac{d^3p}{(2\pi)^3} 
\frac{m}{\beta}\right]}
\eeqa
with $\beta=\sqrt{p_z^2+m^2}$ in the units of the Dirac magnetic moment. 
Note that the magnetization of each Fermi sea has now the opposite
direction and it does not explicitly depend on $U_A$, but the net
magnetization arises by way of the 
exchange splitting of the Fermi sea. Thus we see the ground state holds
ferromagnetism in the presence of $U_A$.

\section{Color magnetic superconductivity}

If FM is realized in quark matter, it might be in the CSC phase.
In this section we discuss a possibility of the coexistence of FM and
CSC, which we call {\it Color magnetic superconductivity} \cite{nak}. 

In passing, it
would be worth mentioning the corresponding situation in condensed
matter physics. Magnetism and superconductivity (SC) have been two major
concepts in condensed matter physics and their interplay has been
repeatedly discussed \cite{MagSup1}. Very recently some materials have been observed to
exhibit the coexistence phase of FM and SC, which properties have not
been fully understood yet; itinerant electrons are responsible to both
phenomena in these materials and one of the important features is 
both phases cease at the same critical pressure \cite{MagSup2}. In our case we shall
see somewhat different features, but the similar aspects as well.

We begin with an OGE-type action:
\beqa
    I_{int}=-g^2\frac{1}{2}\int{\rm d^4}x \int{\rm d^4}y
 \left[\bar{\psi}(x)\gamma^\mu \frac{\lambda_a}{2} \psi(x)\right]
D_{\mu \nu}(x,y)
 \left[\bar{\psi}(y)\gamma^\nu \frac{\lambda_a}{2} \psi(y)\right], 
\eeqa
where $D^{\mu\nu}$ denotes the gluon propagator. 
By way of the mean-field approximation, we have 
\beq
 I_{MF}=\frac{1}{2} \int \frac{{\rm d}^4 p}{(2 \pi)^4} 
                \left( \begin{array}{l} 
                          \bar{\psi}(p)   \\
                          \bar{\psi}_c(p) \\
                       \end{array} \right)^T
                  G^{-1}(p)
                \left( \begin{array}{l} 
                          \psi(p)   \\
                          \psi_c(p) \\
                       \end{array} \right) \\
\label{mfield}
\eeq
in the Nambu-Gorkov formalism.
The inverse quark Green function $G^{-1}(p)$ involves various self-energy 
(mean-field) terms, of which we only keep the color singlet particle-hole 
$V(p)$ and color $\bar 3$ particle-particle ($\Delta$) mean-fields; 
the former is responsible to 
ferromagnetism, while the latter to superconductivity,
\beqa
G^{-1}(p)&=&\left( \begin{array}{cc}
                      \sla{p}-m+\sla{\mu}+V(p) & 
                      \gamma_0 \Delta^\dagger(p) \gamma_0  \\
                      \Delta(p) & 
                      \sla{p}-m-\sla{\mu}+\overline{V}(p) \\
                          \end{array} \right),\nonumber\\
         &=&\left( \begin{array}{cc}
                            G_{11}(p) & G_{12}(p) \\
                            G_{21}(p) & G_{22}(p) \\
                            \end{array} \right)^{-1} \label{fullg2} 
\eeqa
where
\beq
\psi_c(k) = C \bar{\psi}^T(-k),~~~\overline{V} \equiv C V^T C^{-1}. 
\eeq
Taking into account the lowest diagram, we can then write down the 
self-consistent equations for the mean-fields, $V$ and $\Delta$:
\beqa
-V(k)=(-ig)^2 \int \frac{{\rm d}^4p}{i(2\pi)^4} \{-iD^{\mu \nu}(k-p)\} 
      \gamma_\mu \frac{\lambda_\alpha}{2} \{-iG_{11}(p)\} 
      \gamma_\nu \frac{\lambda_\alpha}{2}  
\label{self1}.
 \eeqa
and
\beq
  -\Delta(k)=(-ig)^2 \int \frac{{\rm d}^4p}{i(2 \pi)^4} \{-iD^{\mu \nu}(k-p)\}
               \gamma_\mu \frac{-(\lambda_\alpha)^T}{2}
                 \{-iG_{21}(p) \} 
               \gamma_\nu \frac{\lambda_\alpha}{2}.  
\label{gap1} 
\eeq
\begin{figure}[h]
\begin{center}
\includegraphics[width=5cm,clip]{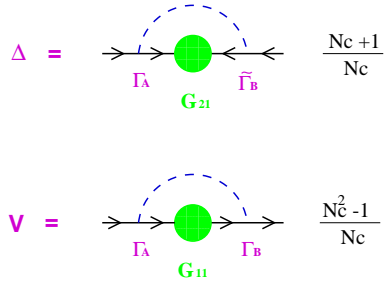}
\end{center}
\caption{Graphical interpretations of the coupled equations (\ref{self1}) and
 (\ref{gap1}) with coefficients in front of R.H.S. given by $N_c$.}
\end{figure}

Applying the Fierz transformation for the Fock exchange energy term (\ref{self1})
we can see that 
there appear the color-singlet scalar, pseudoscalar, vector and axial-vector
self-energies.   
In general we must take into account these self-energies in $V$,
$V=U_s+\gamma_5U_{ps}+\gamma_\mu U_v^\mu+\gamma_\mu\gamma_5U_{av}^\mu$
with the mean-fields $U_i$. Here we retain only $U_s, U_v^0, U_{av}^3$
in $V$ and suppose that others to be vanished.
We shall see this ansatz gives 
self-consistent solutions for Eq.(\ref{self1}) because of axial and reflection
symmetries of the Fermi seas under the zero-range approximation for the
gluon propagator. We furthermore discard the scalar mean-field $U_s$ and 
the time component of the vector mean-field $U_v^0$ for simplicity 
since they are irrelevant for the spin degree of freedom.

According to the above assumptions and considerations 
the mean-field $V$ in Eq.(\ref{fullg2}) renders 
\begin{equation}
V = \gamma_3 \gamma_5 U_A, ~~~U_A\equiv U_{av}^3 , 
\end{equation}
with VA $U_A$.
Then the diagonal component of the Green function $G_{11}(p)$ is written as
\begin{equation}
  G_{11}(p)=\left[ G_A^{-1}-
              \gamma_0 \Delta^\dagger \gamma_0 \tilde{G}_A \Delta \right]^{-1} 
\end{equation}
with
\begin{eqnarray}
   G_A^{-1}(p) &=& \sla{p}-m+\sla{\mu}-\gamma_5 \gamma_3 U_A, \\
  \tilde{G}_A^{-1}(p) &=& \sla{p}-m-\sla{\mu}-\overline{\gamma_5 \gamma_3} U_A,
\end{eqnarray}
where  $\overline{\gamma_5 \gamma_3}=\gamma_5 \gamma_3$ and 
$G_A(p)$ is the Green function with $U_A$  
which is determined self-consistently by way of Eq.~(\ref{self1}).  

Before constructing the gap function $\Delta$, 
we first find the single-particle spectrum 
and their eigenspinors in the absence of $\Delta$, which 
is achieved by diagonalization of the operator $G_A^{-1}$. We have
already known four single-particle energies 
$\epsilon_\pm$ (positive energies) and $-\epsilon_\pm$ (negative energies), 
which are given as 
\begin{eqnarray}
&& \epsilon_{\pm}({\vp}) = 
  \sqrt{{\vp}^2 + U_A^2 + m^2 \pm
             2 U_A \sqrt{m^2 + p_z^2 }}, 
\label{eig}
\end{eqnarray}
and the eigenspinors $\phi_s,~s=\pm $ should satisfy the equation, 
$G_A^{-1}(\epsilon_s, {\bf p})\phi_s=0$. 

Here we take the following ansatz for $\Delta$:  
\begin{eqnarray}
\Delta({\vp})&=&\sum_{s=\pm} \tilde{\Delta}_s({\vp}) B_s({\vp}),\nonumber\\
B_s({\vp})&=&\gamma_0 \phi_{-s}({\vp}) \phi_{s}^\dagger({\vp}).
\label{delta}
\end{eqnarray} 

The structure of the gap
function (\ref{delta}) is then inspired by a physical consideration of a
 quark pair as in the usual BCS theory: 
we consider here the quark pair on each Fermi surface 
with opposite momenta, ${\bf p}$ and $-{\bf p}$ so that they result in a linear combination of
$J^\pi=0^-, 1^-$ (see Fig.~3).
\footnote{Note that this choice is not unique; actually we are now
studying another possibility of quark pair between different Fermi
surfaces \cite{naw}.}
\begin{figure}[h]
\includegraphics[width=6cm,clip]{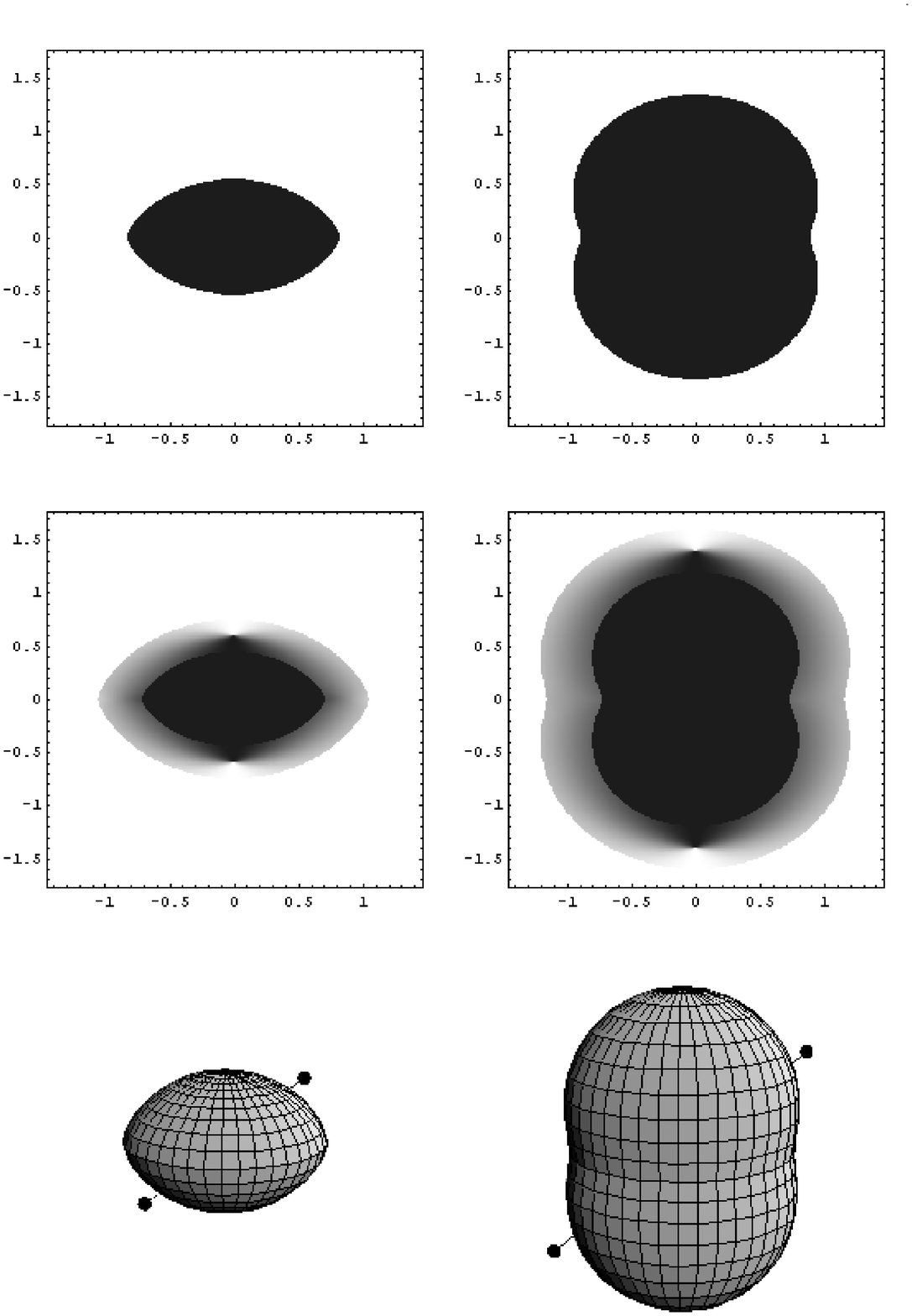}
\includegraphics[width=5cm,clip]{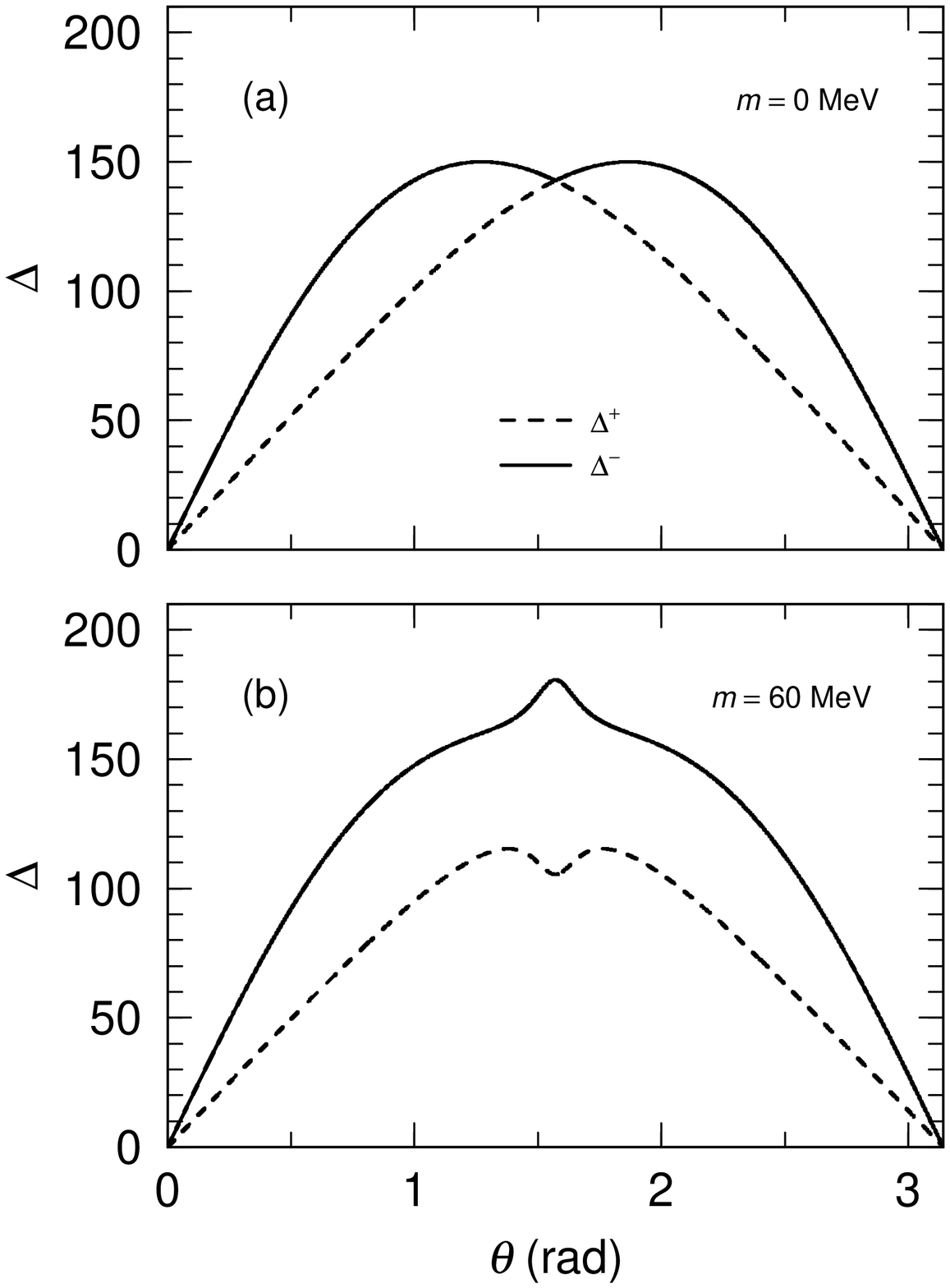}
\sidebyside{\caption{Deformed Fermi seas and the quark pair on each surface. The top
 figures show those in the absence of $\Delta_\pm$ and the middle
 figures  diffusion of the Fermi surfaces in the presence of 
 $\Delta_\pm$. The bottom ones show the quark pairing on the Fermi surfaces.}}
{\caption{Schematic view of the polar-angle dependence of the gap
 functions at the Fermi surface, (a) for $m=0$ and (b) for $m\neq 0$.}}
\end{figure}

 $\tilde\Delta_s$ is still a matrix in the color-flavor space. Since  
the antisymmetric nature of the fermion self-energy imposes a constraint 
on the gap function \cite{CSC1},
\begin{eqnarray}
C \Delta({\vp}) C^{-1}=\Delta^T({-\vp}).
\end{eqnarray}
$\tilde{\Delta}_n({\vp})$ must be a symmetric matrix 
in the spaces of internal degrees of freedom. 
Taking into account the property that the most attractive channel of 
the OGE interaction is  
the color antisymmetric ${\bar 3}$ state, it must be in the flavor singlet 
state.


Thus we can  choose the form of the gap function as 
\beq
\left(\tilde\Delta_s\right)_{\alpha\beta;ij}=\epsilon^{\alpha\beta 3}\epsilon^{ij}\Delta_s
\eeq
for the two-flavor case (2SC), where $\alpha,\beta$ denote the color indices and
$i,j$ the flavor indices. 
Then the quasi-particle spectrum can be obtained by
looking for poles of the diagonal Green function, $G_{11}$: 
\beqa
E_{s}({\vp})&&=\left\{
 \begin{array}{ll}
 \sqrt{(\epsilon_s({\vp})-\mu)^2+|\Delta_s({\vp})|^2} & \mbox{for color 1, 2} \\
 \sqrt{(\epsilon_s({\vp})-\mu)^2}            & \mbox{for color 3} 
 \end{array} 
        \right.  
\label{qusiE}
\eeqa
Note that the quasi-particle energy is independent of color and flavor
in this case, since we have assumed a singlet pair in flavor and color. 

Gathering all these stuffs to put them in the self-consistent
equations, we have the coupled gap equations for $\Delta_s$,
\beqa
 \Delta_{s'}(k,\theta_k)\!=\!\frac{N_c\!+\!1}{2N_c} \tilde{g}^2 
 \!\!\int\!\! \frac{{\rm d}p\, {\rm d}\theta_p}{(2\pi)^2} p^2 \sin\theta_p
   \!\!\sum_s T_{s' s}(k,\theta_k,p,\theta_p) 
   \frac{\Delta_s(p,\theta_p)}{2 E_s(p,\theta_p)}, \label{GAP1}  
\eeqa
and the equation for $U_A$,
\beq
  U_A=-\frac{N_c^2-1}{4N_c^2}\tilde{g}^2 \int \frac{{\rm d}^3 p}{(2\pi)^3} 
\sum_s\left\{\theta(\mu-\epsilon_s(\vp))+2v_s^2({\vp})\right\}
\frac{U_A +s \beta_p}{\epsilon_s(\vp)}, 
\label{UA1} 
\eeq
within the ``contact'' interaction, $\tilde g^2\equiv g^2/\Lambda^2$,
   (see Eq.~(\ref{contact})), where $v_s^2(\vp)$ denotes the momentum distribution of the
   quasi-particles. We find that the expression for $U_A$,
   Eq.~(\ref{UA1}), is nothing but the simple sum of the expectation
   value of the spin operator with the weight of the occupation
   probability of the quasi-particles $v_s^2$ for two colors and the
   step function for remaining one color (cf. (\ref{spol})). 

Carefully analyzing the structure of the function $T_{s's}$ in
   Eq.~(\ref{GAP1}),
 we can easily find that the gap function
   $\Delta_s$ should have the polar angle ($\theta$) dependence on the Fermi
   surface,
\beq
\Delta_s(p^F_s,\theta)=\frac{p^F_s(\theta)\sin \theta}{\mu}
\left( -s\frac{m}{\sqrt{m^2+( p^F_s(\theta) \cos \theta )^2}} R + F \right),
\label{polar}
\eeq
with constants $F$ and $R$ to be determined (see Fig.~4).

As a characteristic feature, both the gap functions have nodes at poles
($\theta=0,\pi$) and take the maximal values at the vicinity of equator
($\theta=\pi/2$), keeping the relation, $\Delta_- \geq \Delta_+
$. This feature is very similar to $^3 P$ pairing in liquid $^3$He or
nuclear matter \cite{leg,NM3P}; actually we can see our pairing function
 Eq.~(\ref{polar}) to exhibit an 
effective $P$ wave nature by a genuine relativistic effect by the Dirac spinors.
Accordingly the quasi-particle distribution is diffused (see Fig.~3)

\subsection{Self-consistent solutions}

Here we demonstrate some numerical results; we replaced the original OGE
by the ``contact'' interaction with the cutoff around the Fermi surface
in the momentum space,
\begin{equation}
D^{\mu\nu}\rightarrow -g^{\mu\nu}/\Lambda^2,~~~
\Delta_s({\bf p}) 
\rightarrow \Delta_s({\bf p})\theta(\delta-|\epsilon_s-\mu|)
\label{contact}
\end{equation}
as in the BCS theory in the weak-coupling limit \cite{PiRi}.




First we show the magnitude of $U_A$ (Fig.~5).
It is seen that the axial-vector mean-field (spin polarization) 
appears above a critical density and
becomes larger as baryon number density gets higher.
Moreover, 
the results for different values of the quark mass show that
spin polarization grows more for the larger quark mass.
This is because a large quark mass gives rise to much difference 
in the Fermi seas of two different ``spin'' states, 
which leads to growth of the exchange energy in the axial-vector channel.
A slight reduction of $U_A$ arises as a result of diffuseness of the
Fermi surface due to $\Delta_s$. As seen in Eq.~(\ref{UA1}), $U_A$ can
be obtained as addition and cancellation of the contributions by two different Fermi seas;
the latter term is more momentum dependent than the former one and
thereby $v_s^2({\bf p})$ enhances the cancellation term (see Fig.~3).

\begin{figure}[h]
 \includegraphics[width=5cm,clip]{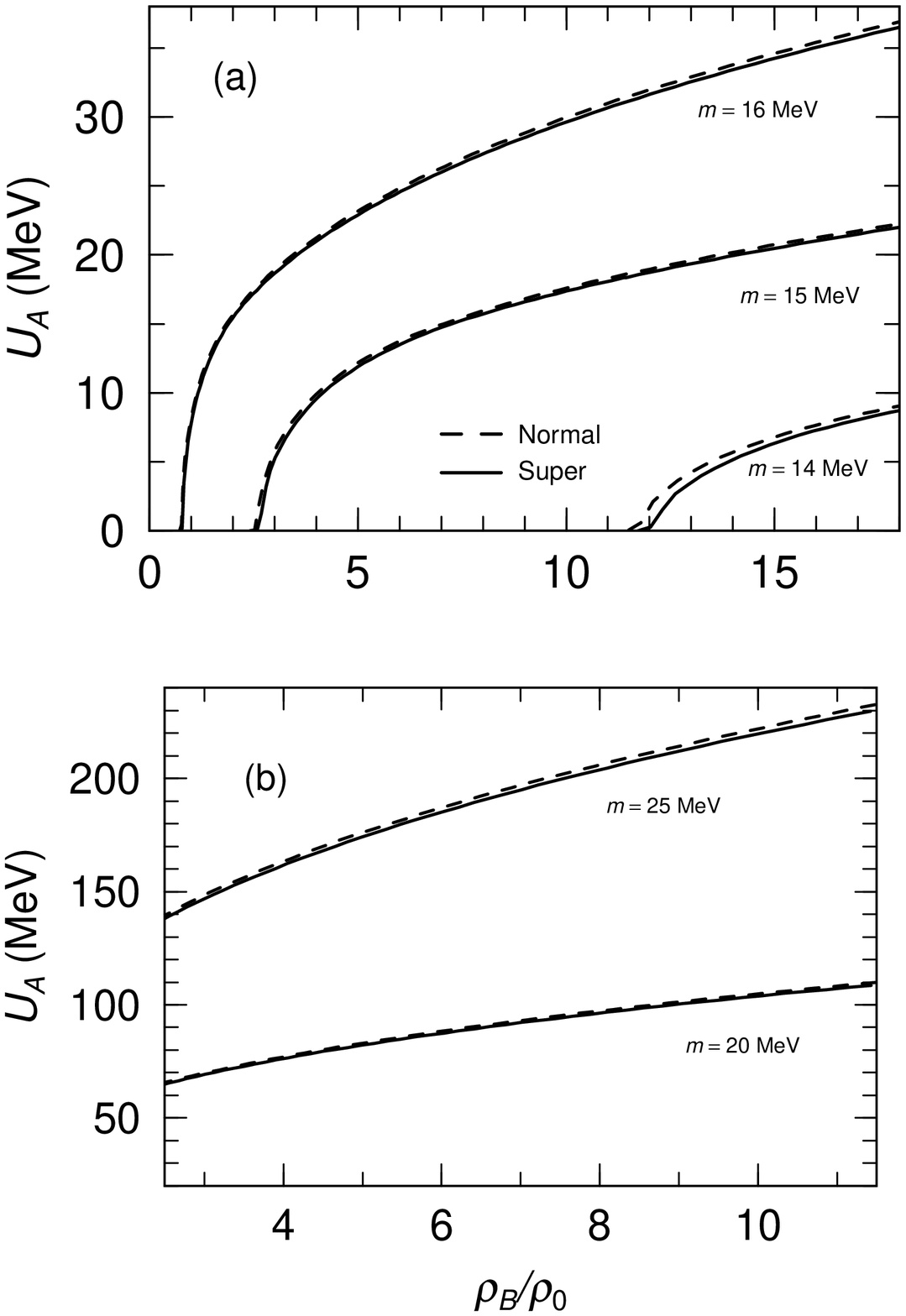}
 \includegraphics[width=5cm,clip]{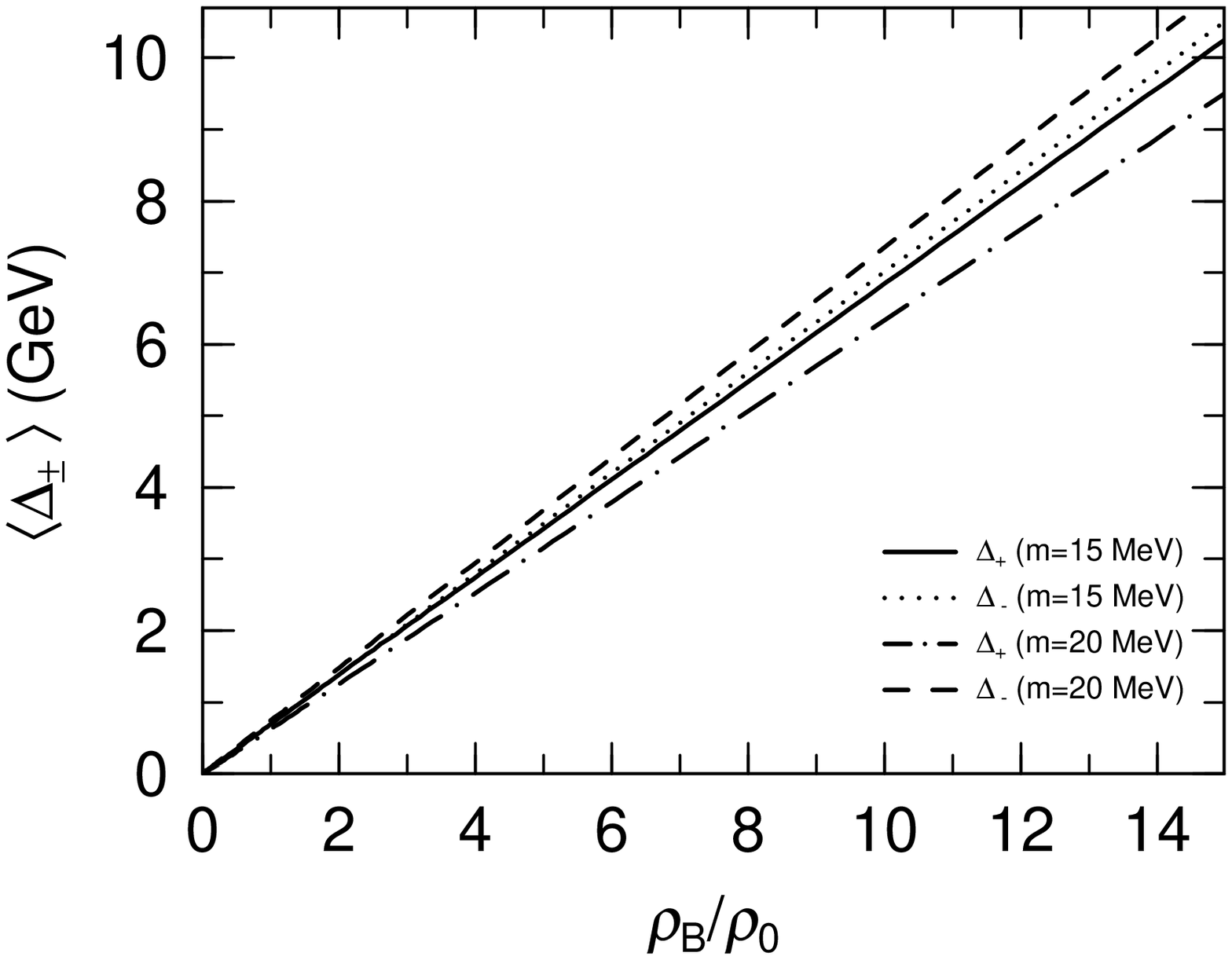}
\sidebyside
{\caption{Axial-vector mean-field (VA) as a function of baryon number density
 $\rho_B$($\rho_0=0.16$fm$^{-3}$). Solid (dashed) lines denote VA in the
 presence (absence) of CSC.}}
{\caption{Mean values of the gap functions, $\Delta_\pm$ and their mass dependence.}}
\end{figure}

Next we show the gap function as a function of $\rho_B$ (Fig.~6). To see the bulk
behavior of the gap function, we use the mean-value with respect to the
polar angle on the Fermi surface,
\begin{equation}
\langle \Delta_\pm \rangle \equiv 
\left( \int_0^\pi {\rm d}\theta \frac{\sin\theta}{2} \Delta_\pm^2 \right)^{1/2}.
\end{equation}
The mean values $\langle \Delta_\pm \rangle$ begin to split with each
other at a density where $U_A$ becomes finite. We'd like to make a
comment here. One may be surprised to see their value of $O({\rm
GeV})$, coming from our parameter choice. However, 
what we'd like to reveal here is
not their realistic values but a possibility of color magnetic
superconductivity and its qualitative features. More realistic study, of
course, is needed by carefully checking our approximations, especially
the contact interaction and the sharp cutoff at the Fermi surface.

With these figures we can say that FM and CSC barely interfere with each
other.


\section{Chiral symmetry and magnetism}

We have seen that the quark mass dependence of ferromagnetism should
be important, while we have treated it as an input parameter. When we
consider the realization of chiral symmetry in QCD, 
the quark mass should be dynamically generated
as a result of the vacuum ``superconductivity''; $q\bar q$ pairs are
condensed in the vacuum. We consider here $SU(2)_L\times SU(2)_R$ symmetry.
Then Lagrangian should be globally invariant under the operation of any group
element with constant parameters, except the  symmetry-breaking
term stemming from the small current mass, $m_c$. Here we'd
like to suggest another mechanism leading to FM in quark matter with
recourse to chiral symmetry. We shall see that FM may be realized 
{\it without accompanying spin polarization}.

Consider the following parameterization for the combination of the 
quark bilinear fields by
introducing the auxiliary fields, $\rho$ and $\theta_i$: 
\beq
\bar\psi\psi+i\gamma_5\mbox{\boldmath$\tau$}
\bar\psi i\gamma_5\mbox{\boldmath$\tau$}\psi
=\rho\exp(i\gamma_5\btau\cdot\btheta).
\eeq
Then it resides on the chiral ``circle'' with ``modulus ''$\rho$ and ``
phase''$\theta_i$, any point on which is equivalent with each other in
the chiral limit, $m_c=0$, and
moved to another point by a chiral transformation. 
We conventionally choose a definite point, 
$\langle {\rm vac}|\rho|{\rm vac}\rangle=f_\pi$ ($f_\pi$: the pion decay
constant) and $\langle {\rm vac}|\theta_i|{\rm vac}\rangle=0$,  for the vacuum, which is
 flavor singlet and parity eigenstate. In the following we
shall see that the phase degree of freedom is related to spin
polarization; that is, the ``phase condensation'' with a non-vanishing value
of $\theta_i$ leads to FM \cite{tatnak}.

Separating the fields $\rho$ and $\theta_i$ into the classical ones 
and fluctuations around them, and discarding any fluctuation the 
mean-field theory proceeds: 
\beq
\rho\rightarrow \langle\rho\rangle[\geq 0],
~~~\theta_i\rightarrow \langle\theta_i\rangle.
\eeq
Assuming the simplest but  nontrivial form of the classical 
chiral angle such that $\theta_3({\bf r})={\bf q\cdot r}, \theta_{1,2}=0$,
we call this set a {\it dual chiral density wave} (DCDW) 
\footnote{Some authors considered 
similar configuration \cite{der} and called a chiral density wave in analogy 
with spin 
density wave (SDW) by Overhauser in condensed-matter physics
\cite{ove}. However, only the scalar density oscillates with finite wave number and 
the pseudo-scalar one is discarded in their ansatz.}
:
\beqa
\langle\bar\psi\psi\rangle&=&\Delta\cos {\bf q\cdot r}\nonumber\\
\langle\bar\psi i\gamma_5\tau_3\psi\rangle&=&\Delta\sin {\bf q\cdot r}.
\label{chiral}
\eeqa
It should be obvious that if $\Delta$ vanishes, the phase degree of freedom
has to become redundant, as seen later. It would be worth mentioning
that similar configuration has been studied in other contexts \cite{dau,kut,tak}.
Note that the configuration in (\ref{chiral}) breaks rotational invariance as
well as translational invariance, but the latter invarianvce is
recovered by an isospin rotation \cite{camp}. 

Taking the Nambu-Jona-Lasinio (NJL) model as a simple but nontrivial 
model \cite{NJL}, we explicitly demonstrate that quark matter 
becomes unstable for a formation of DCDW above a critical 
density; the NJL model has been originally presented 
to demonstrate a realization of chiral symmetry in the vacuum, 
while recently it  been also used as 
an effective model embodying spontaneous breaking of chiral symmetry 
in terms of quark degree of freedom \cite{kle}
\footnote{We can see that the OGE interaction gives the same 
form after the Fierz transformation in the zero-range limit.}
\beq
L_{NJL}=\bar\psi(i\ds{\partial}-m_c)\psi+G[(\bar\psi\psi)^2+(\bar\psi
i\gamma_5\mbox{\boldmath$\tau$}\psi)^2] 
\eeq
Using the mean-field approximation (MFA) with the DCDW configuration, 
we introduce a new quark field $\psi_W$ 
by the  Weinberg transformation \cite{wei},
\beq
\psi_W=\exp[i\gamma_5\tau_3 {\bf q\cdot r}/2 ]\psi,
\label{wein}
\eeq
to get the transformed Lagrangian,
\beq
{\cal L}_{MF}=\bar\psi_W[i\ds{\partial}-M-1/2\gamma_5\tau_3
\ds{q}]\psi_W
-G\Delta^2,
\label{effl}
\eeq
with the dynamically generated mass, $M\equiv -2G\Delta$ and $q^\mu=(0, {\bf q})$. This procedure
embodies translational invariance of the ground state, and shows
that we essentially consider a ``uniform'' problem, while we introduced
the space-dependent mean-fields at the beginning. We briefly summarize
in Table 3 the relation of the transformed frame to the original one. 
Note that the transformed
Lagrangian becomes the same as the familiar form used in discussions of
chiral symmetry realization within the NJL model, except the isovector
and axial-vector coupling term $\gamma_5\tau_3\ds{q}$. We can see the
role of the wave vector $\bf q$ is the same as AV introduced in the
previous sections.

\begin{table}[h]
\caption{Diagram of the Weinberg transformation.}
\centering
\begin{tabular*}{10cm}{@{\extracolsep{\fill}}lcc}
$\langle\bar\psi \psi\rangle\neq 0$~~~&$\Longleftrightarrow$&~~~
$\langle\bar\psi_W \psi_W\rangle=\Delta(\neq 0)$\\
&&\\
$\langle\bar\psi i\gamma_5\tau_3\psi\rangle\neq 0$~~~&&~~~
$\langle\bar \psi_W i\gamma_5\tau_3\psi_W\rangle=0$\\
&&\\
&& ~~~$ q/2\propto \nabla\theta$~~~(``{\rm AV}'')\\
&&\\
{\rm non-uniform} &&~~~~~~~~~{\rm uniform}
\end{tabular*}
\end{table}

The Dirac equation for $\psi_W$ then renders
\beq
\left(i\ds{\partial}-M-1/2\tau_3\gamma_5\ds{q}\right)\psi_W=0.
\label{hf}
\eeq
We can find a 
spatially uniform solution for the quark wave function,
$\psi_W=u_W(p)\exp(i{\bf p\cdot r})$, 
\footnote{This feature is very different from refs.\cite{der}, where the wave 
function is no more plane wave.}
and the energy eigenvalue is given as 
\beq
E_p^{\pm}=\sqrt{E_{p}^{2}+|{\bf q}|^2/4\pm \sqrt{({\bf
p}\cdot{\bf q})^2+M^{2}|{\bf q}|^2}},~~~E_p=(M^2+|{\bf p}|^2)^{1/2}
\label{energy}
\eeq
for positive energy (valence) quarks with different polarizations. For 
negative energy quarks in the Dirac sea, they have a spectrum symmetric with 
respect to null line because of charge conjugation symmetry in the  
Lagrangian (\ref{effl}). 
The single-particle spectrum (\ref{energy})  shows again an analogous
feature to the 
{\it exchange splitting}
between two eigenenergies with different polarizations in the presence of ${\bf q}$; 
hereafter, we choose ${\bf q}//\hat z$, ${\bf q}=(0,0,q),~q\geq 0$, 
without loss of generality. 

Thus each flavor quark shows the same energy
spectrum (\ref{energy}) even in the presence of
the isospin dependent AV and its form is the same 
as in Eq.~(14). However, there is one and important difference from
the previous sections; we have considered the {\it flavor singlet AV},
while we are now considering the {\it isovector AV} here. 
The eigenspinor $u_{W, i}^\pm,~i=u,d$ for each flavor 
satisfies the same Dirac equation for a given energy eigenvalue, 
except the different sign in the AV
term, so that we have 
the different form for each flavor; $u_{W,u}^\pm=u_{W, d}^\mp$.
The mean-value of the spin operator $\Sigma_z=\left(\begin{array}{cc}\sigma_3 & 0\\ 0 & \sigma_3 \end{array}\right)$ 
is then given by 
\beq
\bar s_{z,u}^\pm=\frac{1}{2}u^{\pm\dagger}_{W, u}\Sigma_z u_{W, u}^\pm
=\frac{1}{2}\frac{q/2\pm\beta}{E^\pm_p},
\label{spin}
\eeq
with $\beta=\sqrt{p_z^2+M^2}$ for $u$ quarks. The corresponding
value for $d$ quarks is also given as $\bar s_{z,d}^\pm=-\bar s_{z,u}^\pm$. 
Thus we can see two flavors are oppositely polarized to each other. 
Since the integral of $\bar s_{z, i}^\pm$ over the Fermi seas should be 
finite for $q\neq 0$ for each flavor, 
the spin polarization of each flavor  
is finite but has opposite direction to each other. 
Consequently the total spin polarization or the {\it flavor singlet AV} is always vanished in this case.
However, note that
this result is never conflicted with FM of quark matter by 
considering the magnetization. As we have already noted,
the response of the system to the magnetic field goes through the
magnetization.  Taking into account
the difference of electric charges of two flavors $Q_i$, $Q_u=+2/3 e$ and
$Q_d=-1/3 e$, we can see that each flavor
{\it coherently} contributes to the magnetization, instead.

\subsection{Thermodynamic potential}

The thermodynamic potential is given as
\beqa
\Omega_{\rm total}&=&\gamma\sum_{s=\pm}\int\frac{d^3p}{(2\pi)^3}
(E^s_p-\mu)\theta_s
-\gamma\sum_{s=\pm}\int\frac{d^3p}{(2\pi)^3}E^s_p
+M^2/4G\nonumber\\
&\equiv&\Omega_{\rm val}+\Omega_{\rm vac}+M^2/4G .
\label{therm}
\eeqa
where $\theta_\pm=\theta(\mu-E^\pm_p)$, $\mu$ the chemical potential and 
$\gamma$ is the degeneracy factor $\gamma=N_fN_c$. The first term 
$\Omega_{\rm val}$ is the 
contribution by the valence quarks filled up to the chemical potential, 
while the second term $\Omega_{\rm vac}$ is the vacuum 
contribution that is formally divergent. We shall see both contributions
are {\it indispensable} in our discussion.  
Once $\Omega_{\rm total}$ is properly evaluated, the equations to be solved to 
determine the optimal values of $\Delta$ and $q$ are 
\beq
\frac{\partial\Omega_{\rm total}}{\partial\Delta}
=\frac{\partial\Omega_{\rm total}}{\partial q}=0.
\label{self}
\eeq

Since the NJL model is not renormalizable, we need some regularization procedure 
to get a meaningful finite value for the vacuum contribution.
Consider the sum of the negative energy over the Dirac sea,
\beq
\Omega_{\rm{vac}}=-\gamma\sum_{s=\pm}\int\frac{d^3p}{(2\pi)^3}E^s_p
-\Omega_{\rm ref},
\label{a}
\eeq
where we subtracted an irrelevant constant 
$\Omega_{\rm ref}=-2\gamma\int\frac{d^3p}{(2\pi)^3}E_p$ 
with an arbitrary reference mass $M=M_{\rm ref}$ to make the following 
procedure mathematically well-defined.  Since the energy spectrum
is no more rotation symmetric, we cannot apply the usual momentum cut-off 
regularization (MCOR)
scheme to regularize $\Omega_{\rm{vac}}$. Instead, we adopt the 
proper-time 
regularization (PTR) scheme \cite{sch}. We think this is a most suitable one 
for our purpose, since $\Omega_{\rm{vac}}$ counts the spectrum change under 
the ``external'' axial-vector field.
It has been shown that the vacuum polarization effect   
under the external electromagnetic field can be treated in a gauge 
invariant way, where the energy spectrum is also deformed depending on the 
field strength \cite{sch}. It is also known that the consequences from the NJL model 
are almost regularization-scheme independent \cite{kle}, including the PTR  
scheme.
    
Introducing the proper-time variable $\tau$, we eventually find
\beq
\Omega_{\rm vac}\!=\!\frac{\gamma}{8\pi^{3/2}}\int_0^\infty
\!\!\frac{d\tau}{\tau^{5/2}}
\int^\infty_{-\infty}\!\!\frac{dp_z}{2\pi}\!\left[
e^{-(\sqrt{p_z^2\!+\!M^2}+q/2)^2\tau}
\!+\!e^{-(\sqrt{p_z^2+M^2}\!-\!q/2)^2\tau}\right],
\label{j}
\eeq
except an irrelevant constant $\Omega_{\rm ref}$, which is reduced to the standard formula \cite{kle} 
in the limit $q\rightarrow 0$.

We can easily see, from Eq.~(\ref{j}), that 
the $q$ degree of freedom becomes superfluous and theory must become 
{\it trivial} 
in the limit $m\rightarrow 0$, which is equivalent to $\Delta\rightarrow 0$ 
in the chiral limit: 
all the observables must be independent of $q$. This salient 
feature is consistent with the form of DCDW.
The integral with respect to the proper time $\tau$ is not well defined as 
it is,     
since it is still divergent 
due to the $\tau\sim 0$ contribution.
Regularization proceeds by replacing the lower bound of the integration range 
by $1/\Lambda^2$, which corresponds to the momentum cut-off in the 
MCOR scheme.

For given chemical potential $\mu$, and $M$ and $q$ we can evaluate the valence contribution 
$\Omega_{\rm val}$ using 
Eq.~(\ref{energy}) and write down the general formula 
analytically.
Then the thermodynamic potential can be expressed as
$
\Omega_{\rm val}=\epsilon_{\rm val}(q)-\mu\rho_{\rm val}(q),
$
where $\epsilon_{\rm val}(q)$ and $\rho_{\rm val}(q)$ are the energy density and the quark-number 
density, respectively. They consist of two terms corresponding to the 
two Fermi seas with different polarizations: 
$\epsilon_{\rm val}(q)=\epsilon^-(q)+\epsilon^+(q)$ and 
$\rho_{\rm val}(q)=\rho_{\rm val}^-(q)+\rho_{\rm val}^+(q)$. 
We present some examples about the instability of the usual NJL ground
state with respect to spontaneous generation of DCDW. In the present
calculation chiral symmetry restoration occurs at the first order
in the case without DCDW. 
\footnote{Note that this is not a unique possibility: we may have the second order 
phase transitions for other parameter sets \cite{kle}.}
We can see the NJL
ground state becomes unstable at the critical chemical potential
$\mu_{c1}$, and symmetry restoration is delayed until $\mu_{c2}$ by the
presence of DCDW. This
dragging effect by DCDW is one of the important features.

\begin{figure}[h]
\includegraphics[width=6cm,clip]{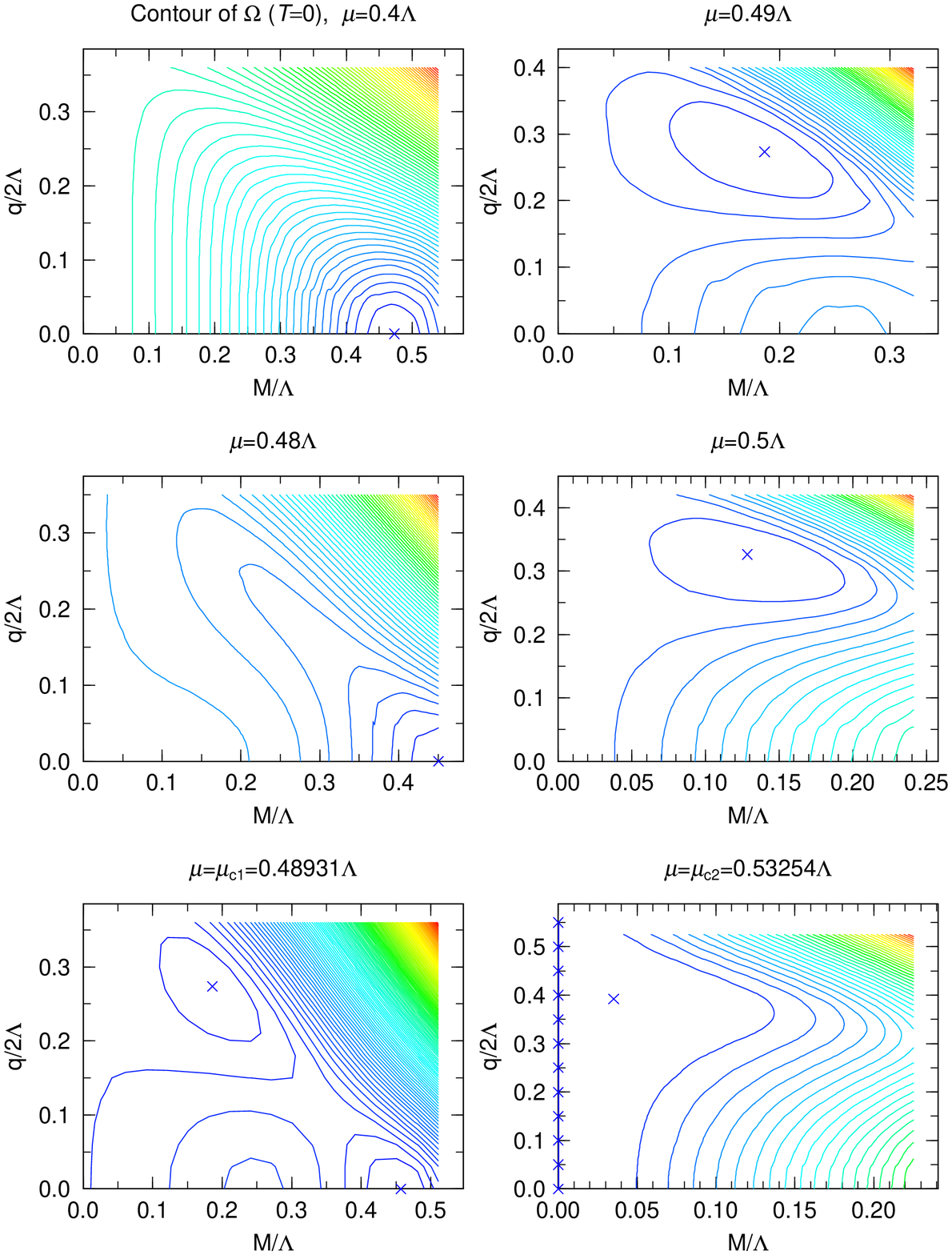}
\includegraphics[width=6cm,clip]{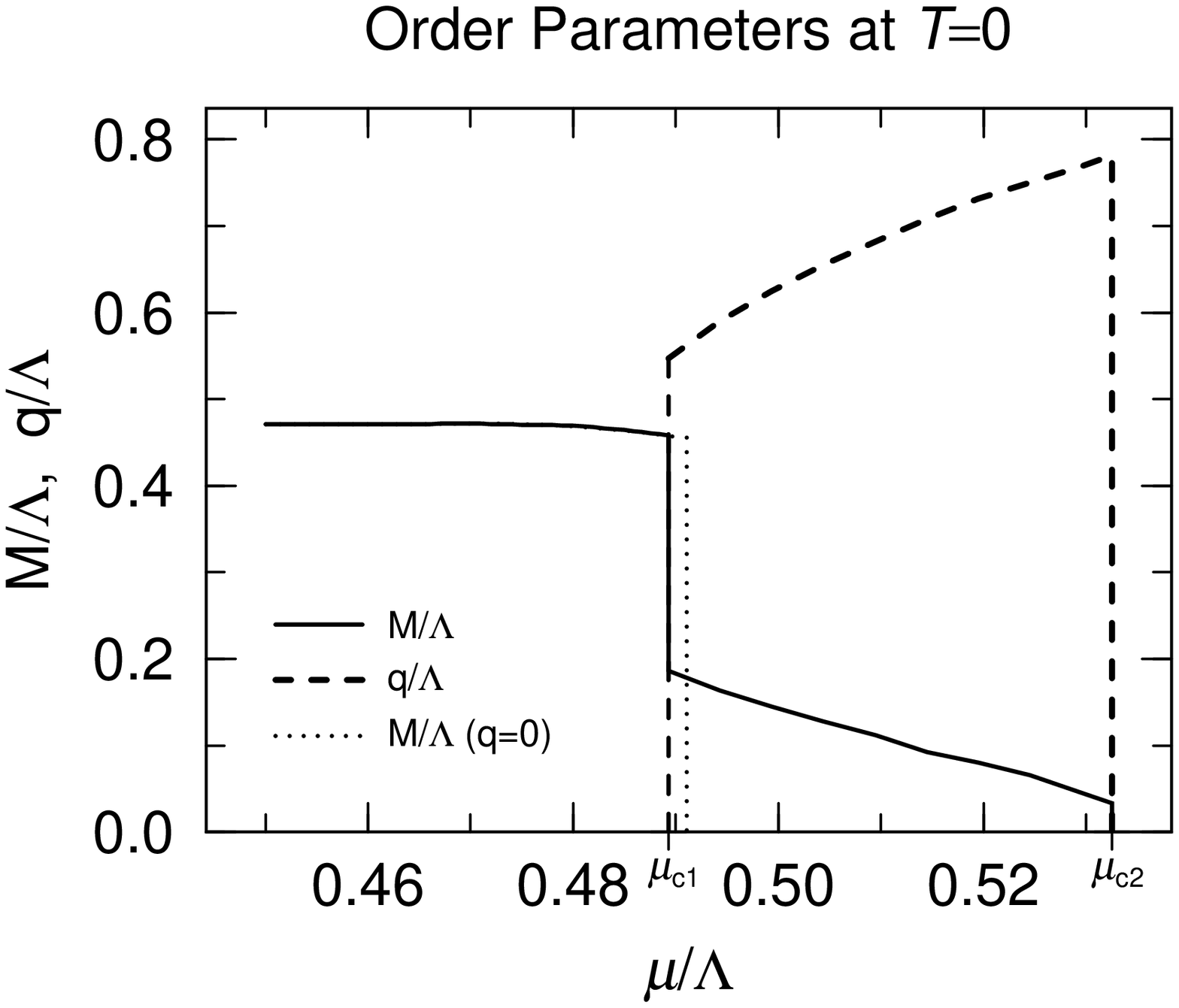}
\sidebyside{\caption{Countour map of the thermodynamic potential in the
 dynamical mass ($M$) - wave number ($q$) plane. The absolute minimum is
 denoted by the cross for given density. We have the first order phase 
transitions in this calculation.}}
{\caption{Density dependence of $M$ and $q$, compared with the usual
 result with $q=0$. There appear two critical chemical potentials; the
 lower one indicates the instability of the ground state for formation of
 DCDW, while the higher one restoration of chiral symmetry.}}
\end{figure}

\section{Summary and Concluding remarks}
  
In this talk we have discussed a magnetic aspect of quark matter based
on QCD. First, we have introduced ``ferromagnetism'' (FM) in QCD, where the
Fock exchange interaction plays an important role. Presence of the
axial-vector mean-field (AV) after the Fierz transformation is essential to
give rise to FM, in the context of self-consistent framework. As one of
the features of the relativistic FM, we have seen that the Fermi sea is
deformed in the presence of AV; the Fermi sea has a prolate shape for
the majority spin particles, while an oblate shape for the minority spin
particles.   

We have then discussed a possibility of color magnetic superconductivity
and seen coexistence of FM and CSC is possible. Our ansatz for quark
pairing shows an effective $P$- wave pair condensation and 
gives a polar angle dependence of the gap function, which looks
similar to liquid $^3$He A - phase. Note that this ansatz is never
unique for color magnetic superconductivity and other types may be also
possible \cite{naw}, where the gap function should show other angle
dependence. In this context recent studies about $S=1$ quark pairing may
be interesting \cite{bub}. 

We have briefly discussed a relation of magnetism to chiral symmetry and
presented an idea, {\it dual chiral density wave} (DCDW), which should
lead to FM. Using, e.g., the NJL model we have demonstrated under what
conditions the ground state becomes unstable for formation of DCDW. We
have found the usual ground state becomes surely unstable at the
critical density and stays in FM between the
first and the second critical densities. 

The FM induced by DCDW has many interesting features different from the
Bloch mechanism. Unfortunately we have not revealed them yet, but it
would be interesting to examine whether DCDW is possible in the CSC phase.  

The symmetry breaking pattern is summarized as follows:
in the condensation of the flavor singlet AV, it violates rotation symmetry,
\beq
O(3)\rightarrow O(2),
\eeq
while the DCDW state does flavor symmetry as well as rotation symmetry,
\beq
O(3)\times SU(2)_V\rightarrow O(2)\times U_{I_3}(1).
\eeq
The latter situation is similar to the neutral pion condensation in
nuclear matter.

It would be important to figure out the low energy excitation modes
(Nambu-Goldstone modes)
built on the ferromagnetic phase. The spin waves are well known in the
Heisenberg model \cite{yoshi}. Then , how about our case \cite{san}?

If quark matter is in the ferromagnetic phase, it may produce the
dipolar magnetic field by their magnetic moment. Since the total magnetic
dipole moment $M_q$ should be simply given as
$M_q=\mu_q\cdot(4\pi/3\cdot r_q^3)n_q$ for the quark sphere with the
quark core radius $r_q$ and the quark number density $n_q$. Then 
the dipolar magnetic field at the star surface $R$ takes the maximal
strength at the poles,
\beq
B_{\rm
max}=\frac{8\pi}{3}\left(\frac{r_q}{R}\right)^3\mu_qn_q=10^{15}[{\rm
G}]\left(\frac{r_q}{R}\right)^3\left(\frac{\mu_q}{\mu_N}\right)
\left(\frac{n_q}{0.1{\rm fm}^{-3}}\right).
\eeq

We have not considered the electromagnetic interaction between quarks
and the induced magnetic field. It would be interesting to see how the
situation changes when we take it into account; 
symmetry restoration \cite{sug} or mixing between magnetic field and
gluon field are among them \cite{alf}.
 
Finally we'd like to give a comment about fluctuations. In this talk we
have completely discarded fluctuations and been only  concerned with the
mean-field. It would be reasonable to study the phase transition, at
least qualitatively. However, we know some fluctuations or correlations
between relevant operators should have some effects even before the
phase transitions. In particular the axial and magnetic susceptibilities in normal quark
matter would be interesting; they might have important consequence,e.g., for 
quark-quark pairing correlation as in $^3$He  superfluidity \cite{leg}.



\begin{acknowledgments}
The present work of T.T. and T.M. is partially supported by the 
Japanese Grant-in-Aid for Scientific Research Fund of the Ministry
of Education, Culture, Sports, Science and Technology (11640272,
13640282), and by the REIMEI Research
Resources of Japan Atomic Energy Research Institute (JAERI). 
\end{acknowledgments}



%




\end{document}